\documentstyle[12pt,aaspp4,flushrt]{article}

\def\be{\begin{equation}}
\def\ee{\end{equation}}

\def\half{{\textstyle{1\over2}}}

\def\bfw{{\bf w}}

\def\omit#1{}
\def\pa{\partial}

\def\bfz{{\bf z}}

\def\bfa{{\bf a}}
\def\bft{{\bf t}}

\def\bfG{{\bf G}}

\def\bfI{{\bf I}}

\begin{document}

\title{Canonical elements for collision orbits}

\author{Scott Tremaine}

\affil{Princeton University Observatory, Peyton Hall, \\
Princeton, NJ~08544-1001, USA}

\begin{abstract}
I derive a set of canonical elements that are useful for collision orbits
(perihelion distance approaching zero at fixed semimajor axis). The
coordinates are the mean anomaly and the two spherical polar angles at
aphelion.
\end{abstract}

\noindent
The purpose of this short note is to derive a set of canonical orbital
elements that are well-behaved for highly eccentric Kepler orbits. More
precisely, I wish to find a set of elements that remain well-defined for
collision orbits (eccentricity $e\to 1$ at fixed semimajor axis $a$). Such
elements may be particularly useful for describing the orbital evolution of
long-period comets.

The Delaunay elements are defined as
\be
L=(M_\star a)^{1/2}, \quad G=L(1-e^2)^{1/2}, \quad H=G\cos i, \quad \ell,
\quad g, \quad h;
\label{eq:delaunay}
\ee
here $M_\star$ is the mass of the star in units where the gravitational
constant is unity; $a$, $e$, and $i$ are the semimajor axis, eccentricity, and
inclination; and $\ell$, $g$ and $h$ are the mean anomaly, argument
of perihelion, and longitude of the ascending node. The vector $\bfG$ is the
specific angular momentum, $G=|\bfG|$ is its magnitude, and
$H=\bfG\cdot\hat\bfz$ is its $z$-component.  In these variables the
Kepler Hamiltonian is
\be
H_K(L)=-{M_\star^2\over 2L^2}.
\label{eq:kepham}
\ee

Any set of six scalar orbital elements is indeterminate for some orbits
(e.g. \cite{ss71}).  For example, Delaunay elements are unsuitable for
circular orbits ($\ell$ and $g$ indeterminate), equatorial orbits ($g$ and $h$
indeterminate), and collision orbits ($i$, $g$ and $h$
indeterminate). Poincar\'e elements are suitable for circular orbits;
eccentric and oblique elements are suitable for circular and for equatorial
orbits; but all of these element sets are indeterminate for collision orbits
(Murray \& Dermott 1999). Herrick (1971) gives an extensive discussion of
universal variables, which are well-defined for all Keplerian orbits including
collision orbits, but Herrick's variables are not canonical.

The orientation of the line of apsides of an orbit can be specified by the
spherical polar coordinates $(\theta_a,\phi_a)$ of the unit vector $\hat\bfa$
pointing from the origin to aphelion, which are given in terms of the elements
$i$, $g$ and $h$ by
\be
\cos\theta_a=-\sin g\sin i, \quad
\sin\theta_a\sin(\phi_a-h)=-\sin g\cos i, \quad 
\sin\theta_a\cos(\phi_a-h)=-\cos g.
\label{eq:defapse}
\ee
The angles $(\theta_a,\phi_a)$ remain well-defined for collision orbits and
thus are attractive candidates for canonical coordinates. I now demonstrate
that the following is a set of canonical elements:
\be
L, \quad \Theta, \quad H, \quad \ell, \quad \theta_a, \quad \phi_a,
\ee
where $\Theta$ is the component of the angular momentum along a line
perpendicular to the $z$-axis and the line of apsides. More specifically, if
$\hat\bfa$ is the unit vector pointing towards aphelion, and we define a unit
vector $\hat\bft\equiv \hat\bfz\times\hat\bfa/\sin\theta_a=
(-\sin\phi_a,\cos\phi_a,0)$, then $\Theta=\bfG\cdot\hat\bft$, thus specifying
the sign of $\Theta$.  The elements $\theta_a$ and $\phi_a$ are indeterminate
for circular orbits, and $\Theta$ is indeterminate if the line of apsides
coincides with the $z$-axis.

Consider a canonical transformation from the Delaunay elements
(\ref{eq:delaunay}) to new momenta and coordinates $(\bfI,\bfw)$, defined by
the mixed-variable generating function
\begin{eqnarray}
S(L,G,H,\bfw)&=& w_1L+(w_3-\half\pi)H-\half\pi G \nonumber \\
&& \qquad \pm G\cos^{-1}{G\cos w_2\over (G^2-H^2)^{1/2}} \mp
H\cos^{-1}{H\cot w_2\over (G^2-H^2)^{1/2}},
\end{eqnarray}
where $0\le w_2\le\pi$ and $0\le w_3<2\pi$. The arguments of the inverse
trigonometric functions have absolute value less than unity if 
\be
\sin w_2>|\cos i|,
\ee 
that is, $w_2$ lies between $\half\pi-i$ and $\half\pi+i$. Then
\be
I_1={\pa S\over\pa w_1}=L, \qquad I_2={\pa S\over\pa w_2}= 
\pm\left(G^2-{H^2\over\sin^2 w_2}\right)^{1/2}, \qquad I_3={\pa S\over\pa
w_3}=H, 
\ee
and
\begin{eqnarray}
\ell & = & {\pa S\over\pa L}=w_1, \qquad g={\pa S\over\pa G}=-\half\pi
\pm\cos^{-1}{G\cos w_2\over(G^2-H^2)^{1/2}}, \nonumber \\
& & \qquad h={\pa S\over \pa H}=
w_3-\half\pi\mp\cos^{-1}{H\cot w_2\over (G^2-H^2)^{1/2}}.
\end{eqnarray}
Thus $-\pi\le g+\half\pi\le\pi$, and the upper sign applies when
$g+\half\pi\ge0$ and the lower when $g+\half\pi< 0$. 

The equations for $g$ and $h$ imply that 
\begin{eqnarray}
\cos w_2=-\sin g\sin i, \quad & &\sin(w_3-h)=\cot i\cot
w_2=-{\sin g\cos i\over\sin w_2}, \nonumber \\
\cos(w_3-h)& = &\mp(1-\cot^2i\cot^2w_2)^{1/2}=-{\cos g\over\sin w_2}.
\label{eq:angdef}
\end{eqnarray}
Comparison with equations (\ref{eq:defapse}) then shows that $w_2=\theta_a$
and $w_3=\phi_a$. It is now straightforward to show that $I_2=\Theta$.

\end{document}